\documentclass[10pt,letterpaper,twocolumn]{article}
\usepackage{ol2}
\usepackage{graphics}
\usepackage[draft]{hyperref}
\usepackage{amsmath}

\begin{document}

\twocolumn[
\title{All-optical delay of images by backward four-wave mixing in metal-nanoparticle composites}

\author{Kwang-Hyon Kim,$^{1,2}$, Anton Husakou$^1$ and Joachim Herrmann$^1$}

\address{$^1$Max Born Institute for Nonlinear Optics and Short Pulse Spectroscopy,
Max-Born-Str. 2a, Berlin D-12489, Germany
\\
$^2$Institute of Lasers, State Academy of Sciences, Unjong District, Pyongyang, DPR Korea}
\begin{abstract} We theoretically study  a novel method for all-optical delay of images based on backward four-wave mixing in composites containing metal nanoparticles. In this approach a delayed phase conjugate probe pulse is generated by the interaction of two counter-propagating pump beams and a non-collinear shaped probe pulse in the nanocomposite.The fractional delay and the reflectivity for the phase-conjugate signal pulses are studied as function of the input pump intensity. It is shown that this scheme can be used for delayed imaging combined with the elimination of optical diffraction. The advantages of this method include minituarized design, tunable wavelength range up to the telecommunication range and wide bandwidth.
\end{abstract}

\ocis{190.5530, 190.5040, 250.5403.}

 ] 

\noindent

Recent progress in the devolopement of slow light devices has paved the way
towards promising applications for storing, switching and processing of
optical signals, increased resolution of spectroscopic interferometers, as
well as in many other fields, e.g. nonlinear optics and quantum optics (see
e.g. \cite{Khurgin}). Recently it has been demonstrated that instead of
single light pulses two- or three-dimensional images may also be slowed and
stored. All-optical methods for delaying images can find 
applications in holography and optical pattern correlation measurements or
all-optical image routers. The ability to slow images and delay them was
demonstrated by using the effect of electromagnetically-induced transparency (EIT) \cite{Camacho(2007),Pugatch, Shuker, Firstenberg, Vudyasetu(2008)} and coupled image
resonators \cite{Tomita(2010),Sultana(2010)}. Applications in this field
require methods for all-optical delay of images at arbitrary wavelenths (in
particular at telecommunication wavelenths), large delay-bandwidth products,
short (picosecond) pulse durations, and design suitable for on-chip
intergration. Recently the authors of this letter 
have proposed a new approach for a slow light device based on the coherent
interaction between pump and probe pulses in composite materials doped with
metal nanoparticles\cite{arXiv}. In such materials near the plasmon resonance the
absorption becomes saturated\cite{KHKim(2010)}. Therefore in a pump-probe regime an absorption
dip in the homogenous plasmonic absorption profile and a steep change of the
effective refraction index near the probe wavelenhth are created, leading to
a significantly lower group velocity of the probe. In a non-collinear
configuration with a TiO$_{2}$ film doped with silver nanorods a large optical
delay with a delay-bandwidth product of more than 60 for picosecond pulses
at 1550 nm has been predicted.

Besides diverse other methods, backward degenerate four-wave mixing (DFWM)
has also been studied for the realization of slow light in photorefractive
crystals \cite{Sturman2(2009),Mathey(2011)}. The main drawback of slow light
by using photorefractive materials is the very large reponse time, leading
to a severely limited bandwidth. Backward DFWM results in phase conjugation 
\cite{He(2002)} which is an important technique for imaging appications. In this Letter, we study an
approach for all-optical delay applicable also for the delay of images based
on composite materials doped with metal nanoparticles (NPs) and backward
DFWM. In this scheme two counter-propagating pump pulses and an initial
non-collinear pulse-shaped probe pulse excite the nanoparticle composite
(see Fig. 1). As the result of the nonlinear interaction, a signal wave which
is phase-conjugate with respect to the probe wave is generated. Due to the
retarded nonlinear response of the metal nanoparticles in the ps-scale time
range an absorption dip and a corresponding steep change of the refraction
index lead to a delay of the signal wave. This process can be used for
delayed imaging.

First, we study the mechanism of slow light in metal nanocomposites by using
backward degenerate four-wave mixing. As presented schematically in Fig. 1
we consider a slab with a metal-NP composite illuminated by two
counter-propagating quasi-cw pump beams $\mathbf{E}_{1}$ and $\mathbf{E}_{2}$
and a non-collinear probe pulse $\mathbf{E}_{pr}$, which has an arbitrary
spatial amplitude and phase distribution. The three beams have the same
central frequency and are assumed to be s-polarized. Then a s-polarized
signal pulse $\mathbf{E}_{s}$ is generated through the DFWM process. Since
the two pump beams propagate along the opposite directions ($\mathbf{k}_{1}=-%
\mathbf{k}_{2})$, and the signal wave is phase conjugate with respect to the
probe ($\mathbf{k}_{3}=-\mathbf{k}_{4})$, the phase-matching condition is
automatically satisfied. A main feature in the optical response of the metal
composite in the spectral range of plasmon resonance is the enhancement of
the local electric field in the vicinity of the NPs. In this work we
consider the excitation by picosecond pulses, therefore the transient
response of the metal NPs is mainly influenced by electron-phonon processes
and determined by the electron thermalization and the cooling of the hot
electrons through the thermal exchange with the lattices in the metal. Using
the two-temperature model, the nonlinear change of the dielectric function of
the metal can be described as \cite{withgriebner} 
\begin{equation}
\Delta \varepsilon _{m}\left( t\right) =\frac{\chi _{m}^{\left( 3\right) }}{%
\tau _{ep}}\int_{-\infty }^{t}\left\vert E\left( t^{\prime }\right)
\right\vert ^{2}e^{-\frac{t-t^{\prime }}{\tau _{ep}}}dt^{\prime },  \label{1}
\end{equation}%
where $\tau _{ep}$ is the electron-phonon coupling time in the range of 1-3
ps \cite{Bigot(2000)}, $\chi _{m}^{\left( 3\right) }=\chi _{m}^{\left(
3\right) }(\omega _{0};\omega _{0},-\omega _{0},\omega _{0})$ is the
inherent third-order nonlinear susceptibility of the metal NPs at the pump
wavelength $\omega _{0}$ and $E(t)$ is local total field at the
nanoparticles enhanced by the plasmon resonance. 
\begin{figure}[t]
\centerline{ \includegraphics[width=4.5cm]{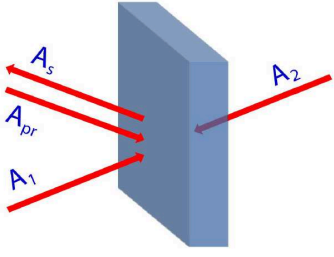}}
\caption[Deleayed images: phase conjugate versus free-space-propagation]{%
Optical configuration for DFWM-based optical delay}
\end{figure}
With the slowly varying amplitudes of each beams $\mathbf{A}_{1}$, $\mathbf{A%
}_{2}$, $\mathbf{A}_{pr}$, and $\mathbf{A}_{s}$ the total electric field is
given by $\mathbf{E}\left( \mathbf{r},t\right) =\mathbf{A}_{1}e^{i\mathbf{k}%
_{1}\mathbf{r}-i\omega t}+\mathbf{A}_{2}e^{i\mathbf{k}_{2}\mathbf{r}-i\omega
t}+\mathbf{A}_{pr}\left( z,t\right) e^{i\mathbf{k}_{3}\mathbf{r}-i\omega t}+%
\mathbf{A}_{s}\left( z,t\right) e^{i\mathbf{k}_{4}\mathbf{r}-i\omega t}+c.c.$. We
assume that the intensity of both quasi-cw pump beams is much higher than
those of the probe and signal waves. Neglecting dispersion and using the slowly-varying envelope
approximation and Eq. (\ref{1}), the propagation of signal and probe pulses 
can be described in the frequency domain as follows: 
\begin{equation}
\left\{ 
\begin{array}{c}
\partial A_{s}\left( \Omega ,z\right) /\partial z=aA_{s}+bA_{pr}^{\ast } \\ 
\partial A_{pr}\left( \Omega ,z\right) /\partial z=bA_{s}^{\ast }+aA_{pr}%
\end{array}%
\right.   \label{2}
\end{equation}%
where $A_{s}\left( \Omega ,z\right) $ and $A_{pr}\left( \Omega ,z\right) $
are amplitudes of signal and probe pulses, $z$ is the coordinate along $\bf{k}_4$,  $%
b=2i\delta A_{1}A_{2}/(1-i\Omega \tau _{ep})$ and 
\begin{equation}
a=i\delta \left( \left\vert A_{1}\right\vert ^{2}+\left\vert
A_{2}\right\vert ^{2}\right) \left( 1+\frac{1}{1+i\Omega \tau _{ep}}\right)
-\alpha /2,  \label{3}
\end{equation}

Here $%
A_{1}$ and $A_{2}$ are the input amplitudes of the two pump waves, $\delta=\omega_{0}\chi _{\mathrm{eff}}^{\left( 3\right) }/2c\mathrm{%
Re}\sqrt{\varepsilon _{\mathrm{eff}}\left( \omega _{0}\right) }\cos \theta$, $\alpha =2\mathrm{Im}\sqrt{\varepsilon _{\mathrm{eff}}\left( \omega
_{0}\right) }\omega _{0}/c$ is the linear absorption coefficient,
 $\theta $ is a half intersecting angle between vectors $\bf{k}_1$ and $\bf{k}_3$, and $\varepsilon _{\mathrm{eff}}\left( \omega _{0}\right) $ is the
linear effective dielectric function of the composite. For spherical
particles with diameters smaller than 10 nm the nonlinear field enhancement
factor $x$ in the frequency damain is given by the implicit relation  (see
Ref. \cite{arXiv})  
\begin{equation*}
x=\frac{3\epsilon _{h}}{\varepsilon _{m0}+2\varepsilon _{h}+\chi
_{m}^{\left( 3\right) }\left\vert xE(\omega _{0})\right\vert ^{2}}
\end{equation*}%
which is solved numerically, and the effective
third-order susceptibility of metal nanocomposite is calculated by $\chi _{%
\mathrm{eff}}^{\left( 3\right) }=f\chi _{m}^{\left( 3\right) }\left\vert
x\right\vert ^{2}x^{2}$ , where $\chi _{m}^{\left( 3\right) }$ is that of
the metal and $f$ is the volume filling factor of NPs. Here $\varepsilon _{m0}$ and $\varepsilon _{h}$
are the permittivities of the metal and the host, respectively and $%
\varepsilon _{\mathrm{eff}}\left( \omega _{0}\right) $ is determined by  the
generalized  Maxwell-Garnett formula including the intensity dependence of $x
$ given by Eq. (4). For spherical NPs with large diameter and nonspherical
NPs we calculate the enhancemant factor $x$ and $\varepsilon _{\mathrm{eff}%
}\left( \omega _{0}\right) $ by using the generalized 
 discrete dipole approximation modified to include nonlinear saturation effects (see
Ref. \cite{KHKim(2010)}).

\begin{figure}[b]
\centerline{ \includegraphics[width=6.5cm]{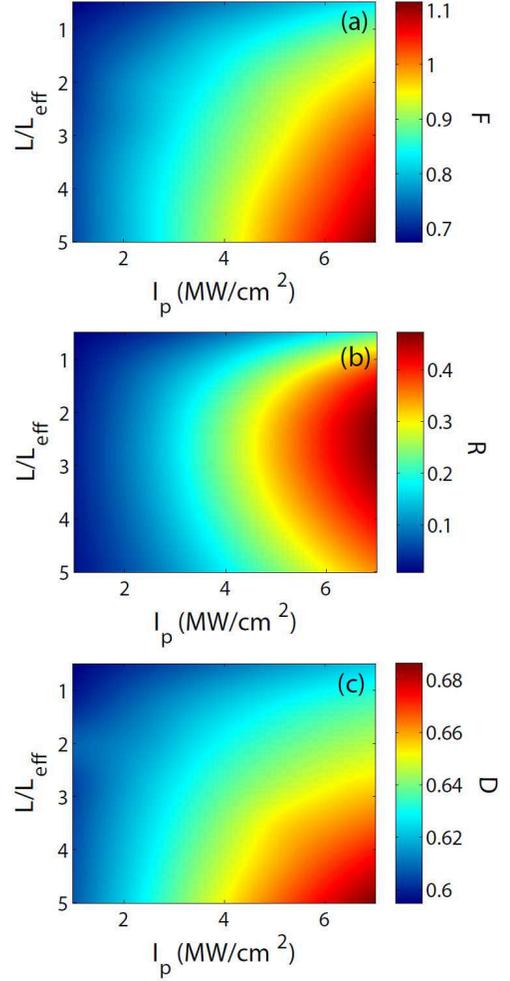}}
\caption{Slow light by DFWM in TiO$_{2}$ layer containing Ag NPs smaller
than 10 nm with filling factor of $10^{-4}$: (a) fractional delay $F$, (b)
conjugate reflectivity $R$, and (c) pulse distortion.}
\end{figure}

In order to characterize the pulse shape change we use the distortion
fuction $D$ of the probe pulse defined by \cite{Shin(2007)} 
\begin{equation}
D=\sqrt{\frac{\int_{-\infty }^{\infty }\left\vert I_{\mathrm{out}}\left(
t+\triangle t\right) -I_{\mathrm{in}}\left( t\right) \right\vert dt}{%
\int_{-\infty }^{\infty }I_{\mathrm{out}}\left( t+\triangle t\right) dt}},
\label{8}
\end{equation}%
where $I_{\mathrm{in}}$ and $I_{\mathrm{out}}$ are normalized input and output probe
pulse intensities, respectively, and $%
\triangle t$ is the peak delay time.

By using the above decribed formalism we now demonstrate the slow light
mechanism by using backward FWM and metal nanocomposites. For a simplified
numerical approach using the Maxwell-Garnett model we consider first a TiO$%
_{2}$ composite containing spherical silver NPs with diameters of 10 nm
which exhibit a plasmon resonance at around 610 nm. In Fig. 2 the fractional
delay (a), the phase conjugate reflectivity (b), and the pulse distortion
(c) are shown as functions of the medium length normalized by the effective
length $L_{\mathrm{eff}}$ $=1/\alpha $ and the pump intensity at 625 nm. The
probe pulse duration is 1 ps. We assume the same intensity of the both
counterpropagating pump beams and an electron-phonon repsponse time of $\tau _{ep}=1$
ps. In this case, the effective length is $L_{\mathrm{eff}}=90.2$ $\mu $m.
Figure 2 shows that by using this configuration the maximum fractional delay
can be in the range of 1 with reflectivity up to more than 0.5 for a pump
intensity lower than 8 MW/cm$^{2}$. The distortion $D$ is about 0.6. 
\begin{figure}[b]
\centerline{\includegraphics[width=7cm]{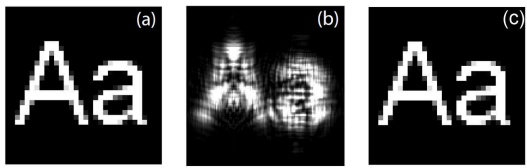}}
\caption{Delayed image (c) by optical phase-conjugation with DFWM and
diffracted image (b) in free-space with the same propagation length at
telecommunication wavelength 1550 nm for a object shown in (a). The nonlinear material is TiO$_{2}$
layer doped with Ag nanorods with a diameter of 22 nm and a length of 64 nm.}
\end{figure}

Let us now study the possibility to use backward FWM in metal
nanocomposites for the purpose of image delay. To demonstrate this mechanism
at telecommunication wavelengths we have chosen a 4.1 $\mu $m-thick TiO$_{2}$ film containing
gold nanorods with a diameter of 22 nm and a length of 64 nm with a filling
factor of 10$^{-3}$. For these nonspherical NPs we use for the numerical
simulations the discrete dipole approximation modified to include nonlinear saturation effects\cite{KHKim(2010)}. The effective nonlinear
susceptibility for gold at the wavelength of 1550 nm is $\chi_3=-1.5\times 10^{-11}$ m$^{2}$V$^{-2}$\cite{value}. Due to field enhancement and the extremely large inherent
nonlinearity of gold at 1550 nm, a pump intensity
of only 30 kW/cm$^{2}$ is sufficient. Figure 3
shows for an object with a total area of 0.32x0.32 mm$^{2}$shown in (a) the delayed image obtained by the backward DFWM (c) as well
as the distorted image after propagation over the same distance of 0.5 mm in
free space without phase conjugation (b). 

One of the main advantages of
this process is that it provides optical phase conjugation \cite{He(2002)}.
A peculiar property of the phase conjugation in the metal-dielctric
composite is that it permits both image reconstruction \textit{and} optical
delay, a combination not commonly found in other schemes for image delay.
 As can be seen from the figure without
phase conjugation diffraction effects are noticeable already after 0.5 mm of
propagation, since the characteristical size of the source features is only
10 nm. However, as shown in Fig. 3(c) phase conjugation reverses the
curvature of the wavefront, resulting in an almost-perfect reconstruction of
the object after propagation over 0.25 mm. 
The
resultant fractional delay of the image in (c) is about 0.75. 

To conculde, we have investigated all-optical delay by backward DFWM in metal
nanocomposite materials. A fractional
delay of a signal pulse has been predicted in the range of unity in both
visible and telecommunication wavelength range. In particular, at
telecommunication wavelength a weak pump intensity in the order of 30 kW/cm$%
^{2}$ can be applied for slowing down of optical pulses. Due to the
optically phase conjugation property of the signal beam with regard to the
probe, all-optical delay of images can be realized. The advantage of the
proposed technique comprises miniaturized design,
wide bandwidth reaching up to THz, and control of
the central wavelength by changing the sizes and shapes of
metal nanoparticles. The optical delay can be further
increased by a thicker nanocomposite with lateral pump incidence as
proposed in \cite{arXiv} or by several parallel composite slabs.

\end{document}